\documentclass[10pt]{iopart}
\usepackage{iopams}
\usepackage{graphicx}

\begin{document}

\title{Laser driven self-assembly of shape-controlled potassium nanoparticles in porous glass}

\author{L. Marmugi$^{1,2}$, E Mariotti$^{1}$, A. Burchianti$^{3}$, S. Veronesi$^{4}$, L. Moi$^{1}$, C. Marinelli$^{1,2}$}

\address{$^{1}$Department of Physical Sciences, Earth and Environment and CNISM, University of Siena, via Roma 56, 53100 Siena (Italy)}
\address{$^{2}$Istituto Nazionale di Ottica-CNR, uos Pisa, via G. Moruzzi 1, 56124 Pisa (Italy)}
\address{$^{3}$Istituto Nazionale di Ottica-CNR and European Laboratory for Non Linear Spectroscopy (LENS), University of Florence, Physics Department, via N. Carrara 1, 50019, Sesto Fiorentino (Italy)}
\address{$^{4}$National Enterprise for nanoScience and nanoTechnology (NEST)-Istituto Nanoscienze CNR, Piazza San Silvestro 12, 56127 Pisa (Italy)}

\ead{carmela.marinelli@unisi.it}

\begin{abstract}
We observe growth of shape-controlled potassium nanoparticles inside a random network of glass nanopores, exposed to low-power laser radiation.  Visible laser light plays a dual role: it increases the desorption probability of potassium atoms from the inner glass walls and induces the self-assembly of metastable metallic nanoparticles along the nanopores. By probing the sample transparency and the atomic light-induced desorption flux into the vapour phase, the dynamics of both cluster formation/evaporation and atomic photo-desorption processes are characterized. Results indicate that laser light not only increases the number of nanoparticles embedded in the glass matrix but also influences their structural properties. By properly choosing the laser frequency and the illumination time, we demonstrate that it is possible to tailor the nanoparticles' shape distribution. Furthermore, a deep connection between the macroscopic behaviour of atomic desorption and light-assisted cluster formation is observed. Our results suggest new perspectives for the study of atom/surface interaction as well as an effective tool for the light-controlled reversible growth of nanostructures.
\end{abstract}


\vspace{2pc}

\maketitle

Atomic adsorption and desorption have a major influence on surface processes and related transport phenomena \cite{mann2009, villalba2010, lenzi2010}. Their control provides a tool for driving the evolution of nanoscale systems, where the atomic mobility is strongly affected by the interaction with the substrate \cite{brault2009, villalba2010, chang2013}. The atomic transport in adsorbing porous materials is governed by the adsoprtion/desorption events at the pore's surface, where atomic nanoaggregates are assembled as a consequence of atomic surface diffusion and nucleation \cite{villalba2010, burchianti2006, burchianti2008, bhagwat2009}. Many experiments involving different adsorbates and substrates demonstrated that light can influence the formation of nanostructures \cite{zeng2014, macdonald2002, callegari2003, fedotov2003, soares2005, srivastav2010}. 

In particular, in porous materials loaded with alkali-metal atoms, it has been proved that atomic photo-desorption \cite{burchianti2004} plays a key role in cluster growth \cite{burchianti2006, burchiantiepjd}. Upon visible illumination, the atomic desorption probability from atomic layers covering the pore's surface increases; consequently also the atomic diffusion coefficient, which is proportional to the desorbing rate, rises. The atomic motion can be thus modelled as a random sequence of free flights between the pore walls, interrupted by adsorption events at the nanopore surface \cite{villalba2010}.  As an atom sticks to the surface, it can be either desorbed again - with a light-enhanced probability - or can be trapped at surface defects where metastable metallic nanoparticles (NPs), characterized by defined surface plasmon bands, are formed \cite{burchianti2006, burchianti2008}.

In this work, we use low-power visible laser irradiation to induce atomic desorption and, thus, the formation of K NPs inside a nanoporous glass matrix. Our results indicate that light, as a consequence of the photo-enhanced atomic mobility, not only increases the atomic concentration at the seed clusters sited on the porous walls \cite{burchianti2006}, but also drives the subsequent process of cluster growth. Indeed, the shape distribution of the light-grown NPs depends on the frequency of the desorbing light as well as on the illumination time. This implies that light acts both as a maker and as a shaper. Furthermore, since the light-grown NPs evaporate with a characteristic time which depends on their shape, the time spent by an atom in a cluster before resuming its random walk along the pore is in turn affected by the light. Therefore, light drives the overall atomic diffusion-reaction dynamics inside the porous matrix.

Unlike the conventional NPs production techniques, relying mainly on chemical \cite{chen2010} and photochemical approaches \cite{zeng2014} or laser ablation in liquids \cite{desarkar2013, popovic2013}, 
the method presented here is based on the reversible light-induced modifications of the atomic adsorption/desorption and cluster nucleation rates. This approach allows for the observation and control of the dynamics of the NPs' formation/evaporation processes. Moreover, it provides the possibility to retrieve the initial conditions of the system, both by intrinsic relaxation and forced resonant evaporation \cite{burchiantiepjd}. Our results, therefore, disclose the path to the development of effective all-optical techniques for the self-assembly of nanoparticles and address the possibility to optically control the surface interactions and the atomic transport in porous media.

\begin{figure*}[htb]
\begin{center}
\includegraphics[width=\linewidth]{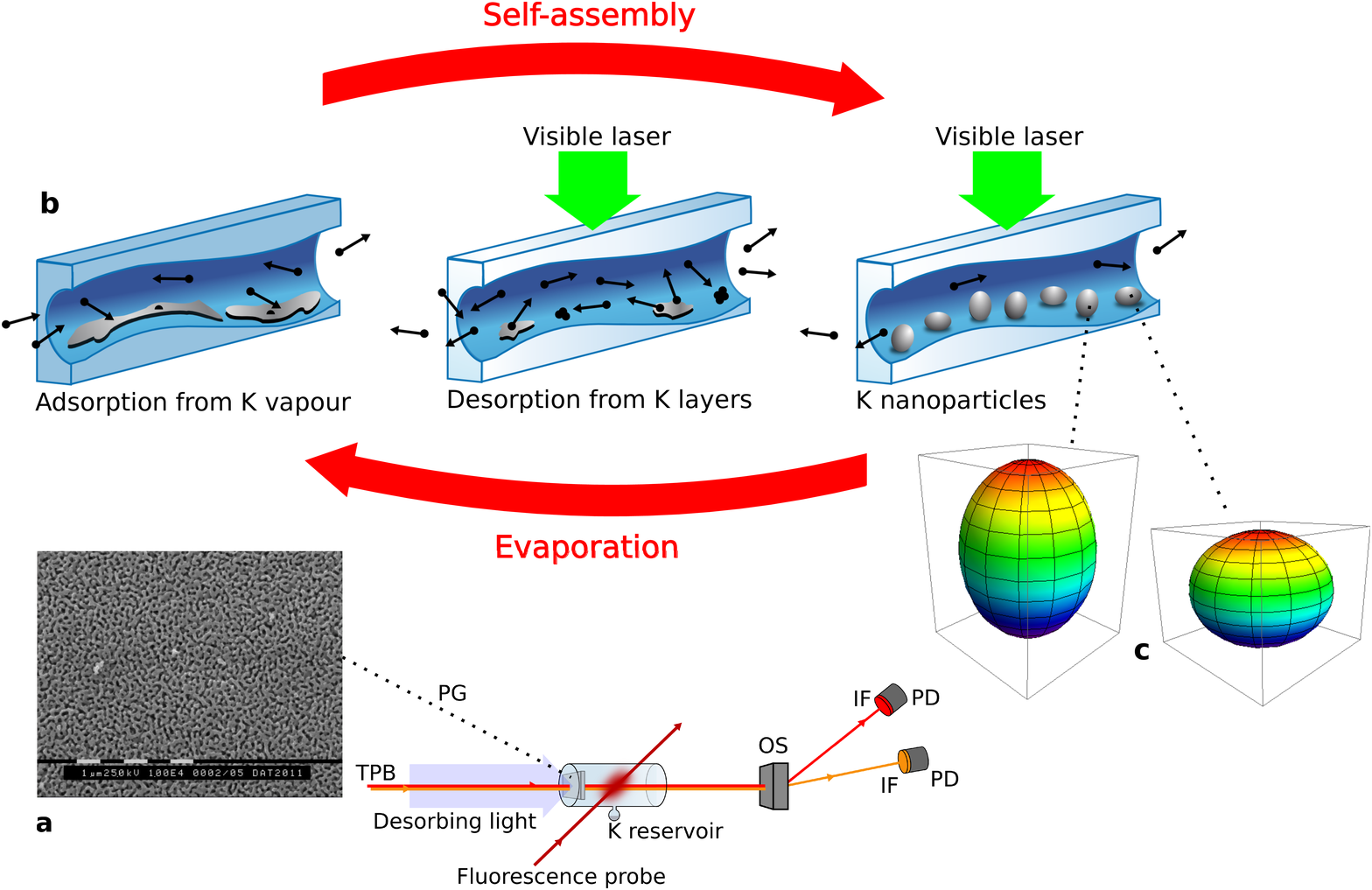}
\caption{Nanoparticles light-driven reversible self-assembly in porous glass. \textbf{a}: Setup for the dynamic PG optical response and REM image of the sample surface (VitraBio GmbH). TPB: transmission probe beams (780 and 850 nm), PG: porous glass sample, OS: optical system, IF: interference filter, PD: photodiode. \textbf{b}: Cross-section of a nanopore showing light-driven reversible assembly of K nanoparticles process. Sketch is not to scale. \textbf{c}: 3D reconstruction of the shape of light-grown K nanoparticles; details are given in the text.}\label{fig:apparatusnp}
\end{center}
\end{figure*}

We use as a host matrix for K atoms and NPs a porous glass (PG) plate (SiO$_{2}$$>$96$\%$), characterised by an interconnected network of randomly oriented pores with a mean diameter of 20 nm. The PG is shaped in plates of 18$\times$15 mm$^{2}$ and an external thickness of 1 mm; it is fixed inside a cylindrical sealed Pyrex cell (24 mm diameter, 80 mm length).  The glass pores are loaded by the continuous exposure to K vapour generated, at room temperature, by a solid reservoir distilled into the cell body. At the equilibrium, the PG plate appears as a semi-transparent, whitish surface due to light scattering provided by the external surface, so no direct imaging of the inner pores can be obtained by optical methods. Sample characterisation is hence performed by Reflection Electron Microscopy (REM) (Fig. \ref{fig:apparatusnp}.a) and porosimetry with mercury intrusion (courtesy of VitraBio GmbH): the internal free volume is 55$\%$, with a total inner surface of about 31 m$^{2}$.

The formation of K NPs under laser irradiation is investigated by measuring the sample optical absorbance with a VIS-NIR spectrophotometer. The PG is exposed to laser light at 405 nm or at 532 nm, generated by a laser diode and by a CW frequency doubled Nd:YAG laser, respectively. An absorbance spectrum is acquired before illumination and is exploited as a reference in order to isolate the contribution of the light-grown nanostructures.

In order to fully characterize the dynamics of the system, the number of the desorbed atoms into the cell volume is measured during and after laser illumination (Fig. \ref{fig:apparatusnp}). The atomic vapour is excited by an extended cavity diode laser (ECDL) tuned on the D$_{1}$ transition at 770.1 nm; the laser-induced atomic fluorescence is collected by a photomultiplier. At the same time, the PG transparency is probed by two free-running laser diodes with emission at 780 nm and 850 nm (Transmission Probe Beams, TPBs). These wavelengths are resonant to the localized surface plasmons at the light-grown NPs surface. The optical paths of the two beams are carefully overlapped inside the sample. The TPBs are attenuated below 4 mW/cm$^{2}$ to prevent local heating. After crossing the porous sample, each beam is spatially separated by an optical system and detected by amplified photodiodes, equipped with dedicated 10 nm pass-band interference filters.  In this configuration, the presence of NPs with absorption bands resonant with the TPBs is revealed by a continuous decrease of the corresponding transmission signals, while a reduction of the clusters number causes a progressive restoration of the signal levels. The sample illumination is provided by the same 405 and 532 nm laser sources exploited for the absorbance measurements.

The absorbance at the equilibrium in the dark indicates the presence in the PG of a small number of  K aggregates with a very broad and almost indistinct size and shape distribution. Such structures spontaneously self-assemble under equilibrium conditions and their lifetime is significantly longer than the typical experimental time scales (some hours). 

\begin{figure}[h]
\begin{center}
\includegraphics[width=\columnwidth]{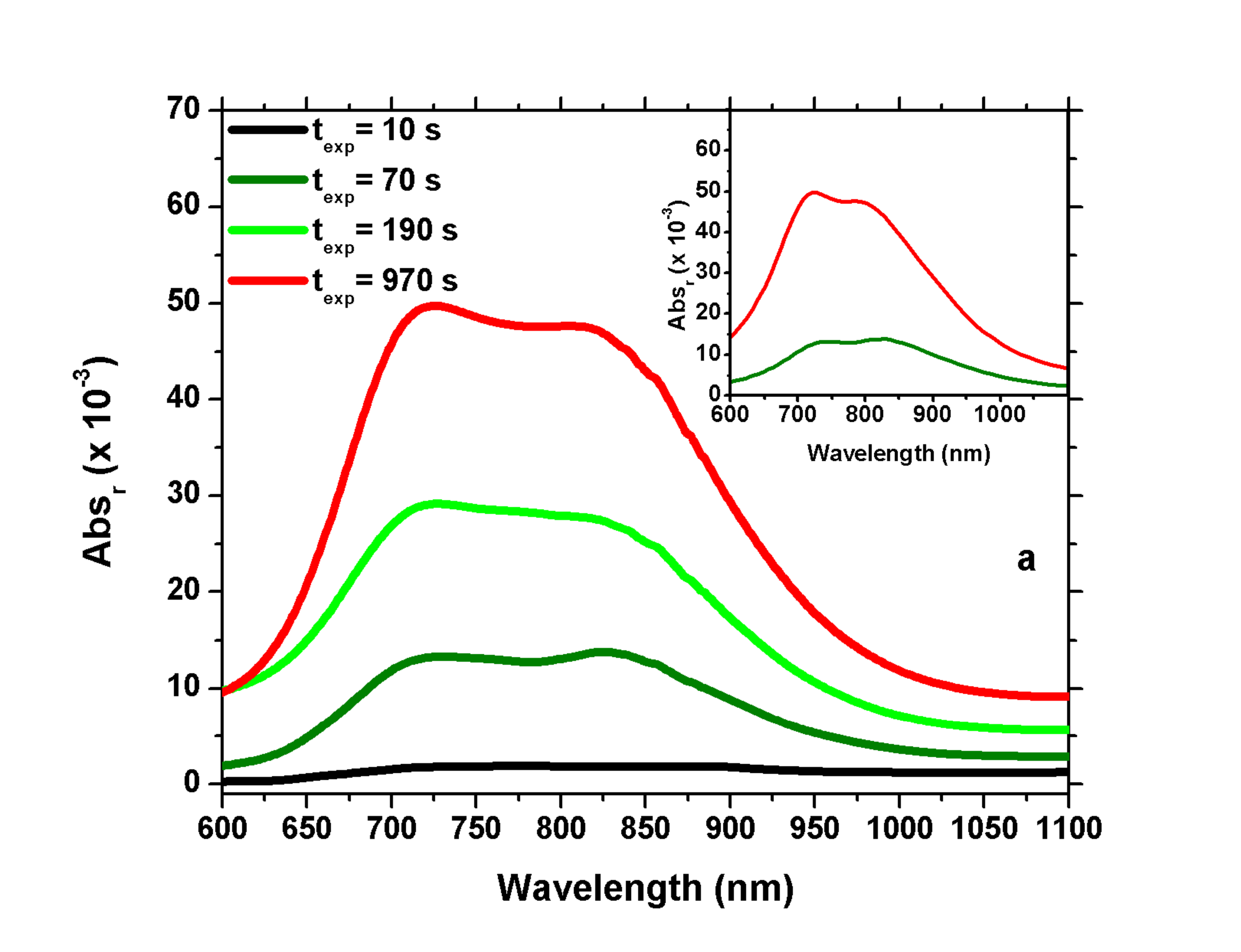}\\
\includegraphics[width=\columnwidth]{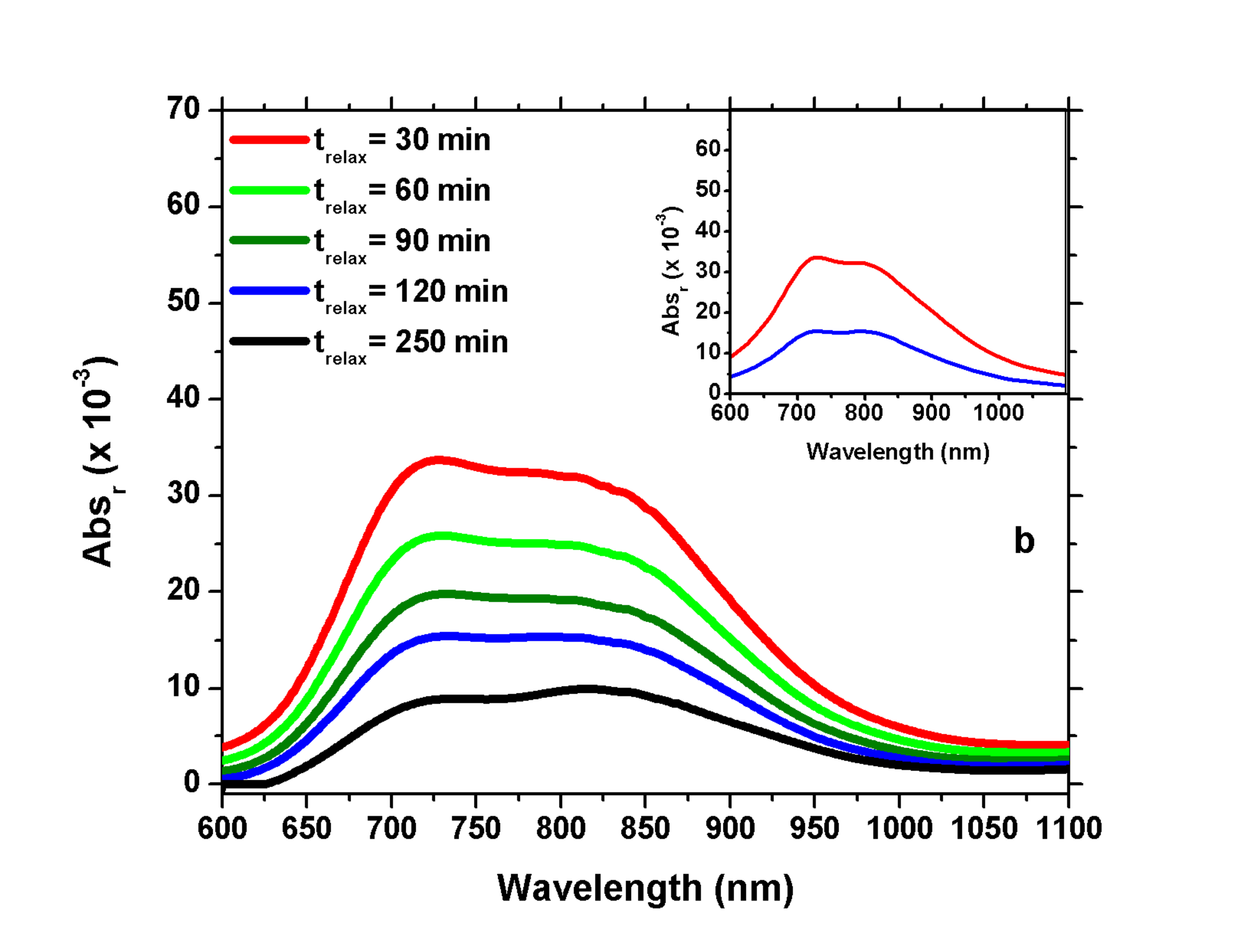}
\caption{\textbf{a}: Referenced sample absorbance after exposure to 350 mW/cm$^{2}$ at 532 nm. Inset: examples of numerical fitting according to the Gans theory. \textbf{b}: Referenced sample absorbance during relaxation in the dark after exposure to 350 mW/cm$^{2}$ at 532 nm for 970 s.  Inset: examples of numerical fitting according to the Gans theory.}\label{fig:greenspectrum}
\end{center}
\end{figure}

The NPs' signatures become more evident after illumination with 532 nm laser light, as shown in Fig. \ref{fig:greenspectrum}.a: a significant amount of new NPs is formed inside the PG sample as a consequence of atomic photodesorption and the following cluster growth. Even if a light-induced increase of the absorbing structures with almost indistinct size and geometry distributions is observed, the most relevant feature of the sample absorbance is the presence of two distinct maxima, centred at 730 nm and 830 nm (\ref{fig:greenspectrum}).a. Such peaks are produced by localized surface plasmons, excited by resonant light at the interface between the K nanoaggregates and the silica substrate \cite{burchianti2006, klempt2006}. Interestingly, the shape of the resonance bands changes by increasing the exposure time: the 730 nm peak progressively overwhelms the 830 nm one.

After exposure to the 532 nm radiation for 970 s, the sample is left in the dark and its absorbance is measured at regular intervals (Fig. \ref{fig:greenspectrum}.b). A general decrease of the absorbance level is observed, demonstrating the intrinsic instability of the light-assembled nanoparticles. This implies the reversibility of the NP production assisted by light, which can be speeded up by resonant light \cite{burchianti2006, burchianti2008, bagratashvili2013}. Moreover, during the spontaneous relaxation in the dark, the shape of the resonance band changes: the 730 nm peak decreases faster than the 830 nm one.

These results are analyzed in the framework of Gans theory \cite{burchiantiepjd}, which provides the optical response of an ensemble of randomly oriented spheroids excited by wavelengths significantly larger than the average size of the nanoparticle. In this way, the random orientation of the pores, which in turn affects the orientation of the NPs, can be effectively taken into account. The extinction cross-section $\sigma_{ext}$ is then calculated as \cite{link1999}:

\begin{equation}
\sigma_{ext}=V \frac{\omega}{3 c} \varepsilon_{m}^{3/2} \sum_{j} \frac{ (1/L_{j}^{2}) \varepsilon_{2}(\omega)}{\left( \varepsilon_{1} (\omega) + \frac{1-L_{j}}{L_{j}}\varepsilon_{m}\right)^{2} + \varepsilon_{2}^{2}(\omega)} \mbox{ ,}\label{eqn:gans}
\end{equation}

where V is the NP volume and $\omega$ the angular frequency of the incoming light. The K dielectric function is $\varepsilon (\omega) \equiv \varepsilon_{1}(\omega) + i \varepsilon_{2}(\omega)$, while the embedding medium response is parametrized by the averaged dielectric constant $\varepsilon_{m}$. The characteristic lengths $L_{j}$ are the depolarising factors for the three symmetry axes of the spheroid (j = a, b, c). More details can be found in \cite{burchiantiepjd} and references therein. It is worth underlining that Eq. \ref{eqn:gans} does not take into account cluster-cluster interaction; moreover, it is based on a rough approximation of the cluster-embedding medium interaction by means of the averaged $\varepsilon_{m}$: interaction causes a red-shift of the actual optical spectrum of the NPs, which is compensated by adding an offset to the calculated spectra.

On the basis of the simulation results, the absorbance peaks are attributed to K spheroids with an average radius of 2 nm nested along the pores. The 830 nm plasmon band is mainly produced by oblates with an axis ratio of 1.25, while the one at 730 nm is mainly due to prolates with an axis ratio of 0.75 (Fig. \ref{fig:apparatusnp}.c).

\begin{figure}[h]
\begin{center}
\includegraphics[width=\columnwidth]{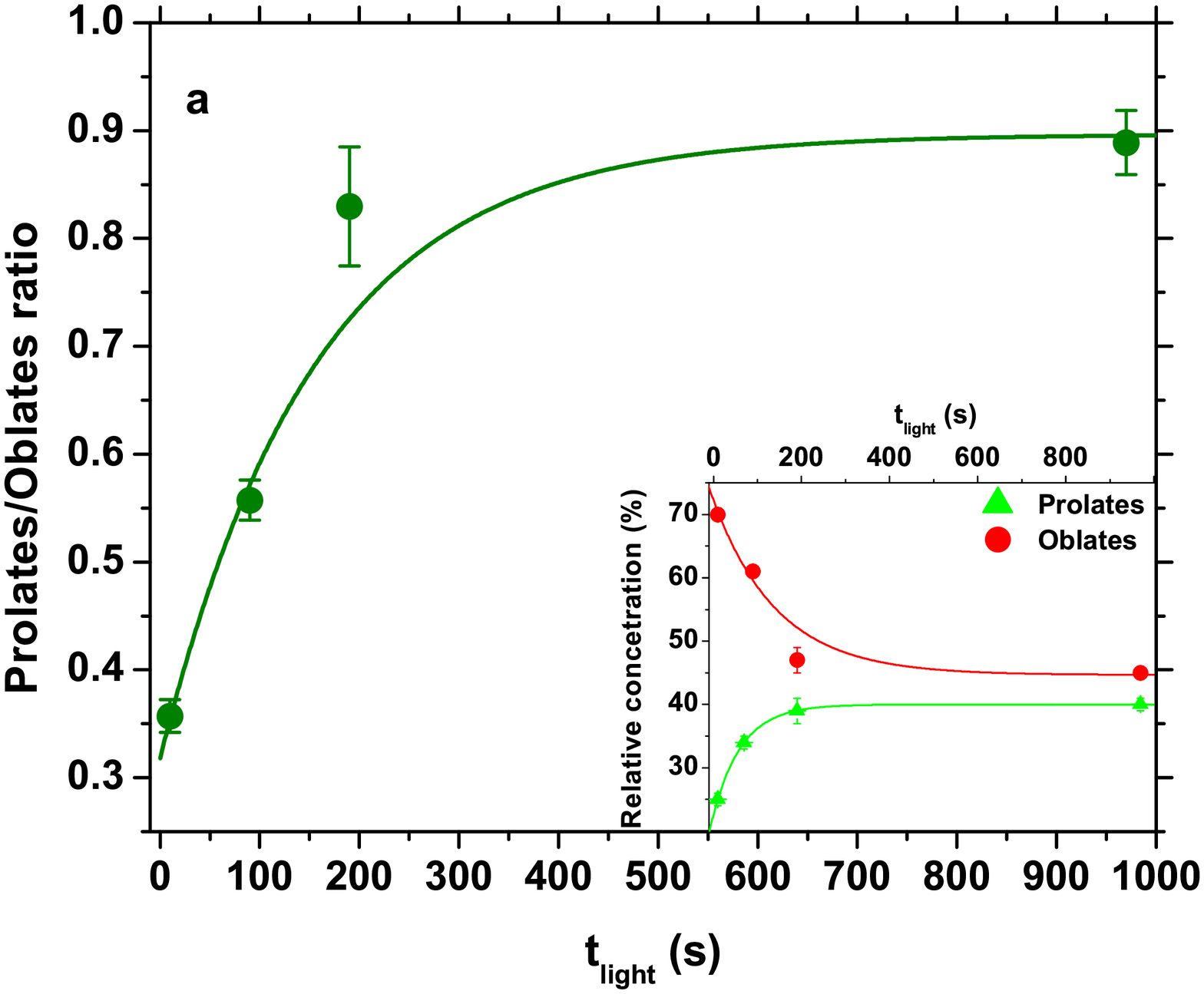}\\
\includegraphics[width=\columnwidth]{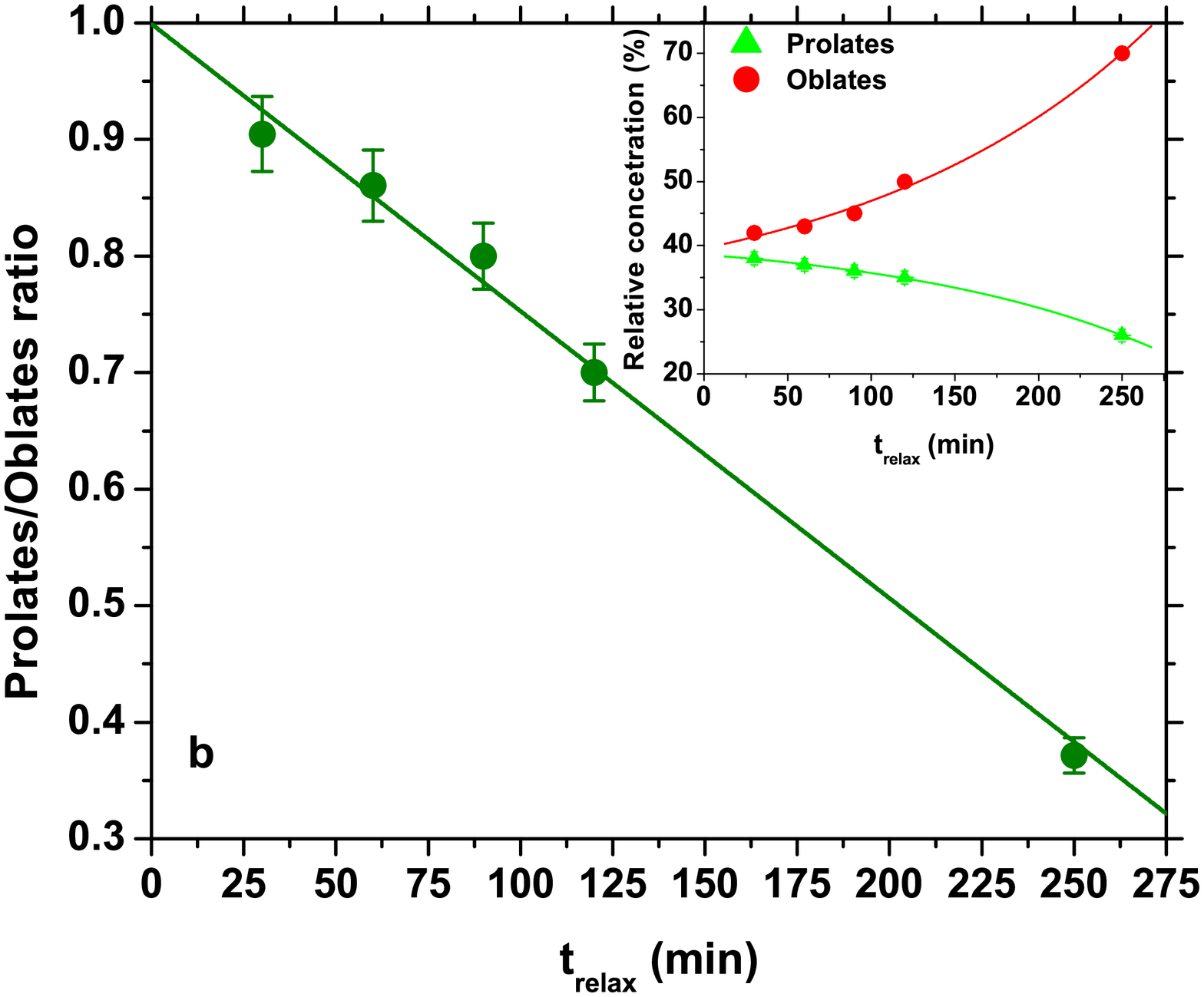}
\caption{Ratio between the number of oblate and prolate NPs as a function of time for green illumination according to the Gans theory fit results. \textbf{a}: 350 mW/cm$^{2}$ at 532 nm. \textbf{b}: relaxation in the dark. Lines are guides to the eye.}\label{fig:greenperc}
\end{center}
\end{figure}

The relative composition of the NPs population can be thus extrapolated (Fig. \ref{fig:greenperc}). While the total number of spheroids is increasing upon illumination, the geometry distribution of light-grown nanoparticles is progressively shifted towards prolates. Only after 10 s, oblates are almost 3-times more frequent than prolates; however, with longer illumination times, the prolates relative concentration increases and equals the concentration of oblates. 

In light of these experimental evidences, we argue that the green laser irradiation produces a progressive modification of the adsorbate/substrate interactions which favours the formation of NPs with a smaller surface contact. It is worth noting that this effect is opposed to the equilibrium nucleation processes, when atoms move spontaneously towards less curve surfaces, that have a lower chemical potential \cite{combe2000}. In this sense, the NPs self-assembly is driven by light: the new nanostructures are far from the thermodynamical equilibrium, contrary to the usual situation for nanoscale self-assembly \cite{grzelczak2010}.

The behaviour observed during relaxation in the dark confirms the intrinsic full reversibility of the process: while the number of NPs is reducing because of their metastability, the oblates dominion is spontaneously restored. This indicates that the prolate nanoaggregates have a higher instability than the oblate ones. The relative concentration of prolates and oblates after 250 min in the dark is compatible with the one obtained after 10 s illumination at 532 nm: the decay process is thus significantly slower than the growth; nevertheless, the initial conditions of the system are spontaneously restored in a few hours.

\begin{figure}[h]
\begin{center}
\includegraphics[width=\columnwidth]{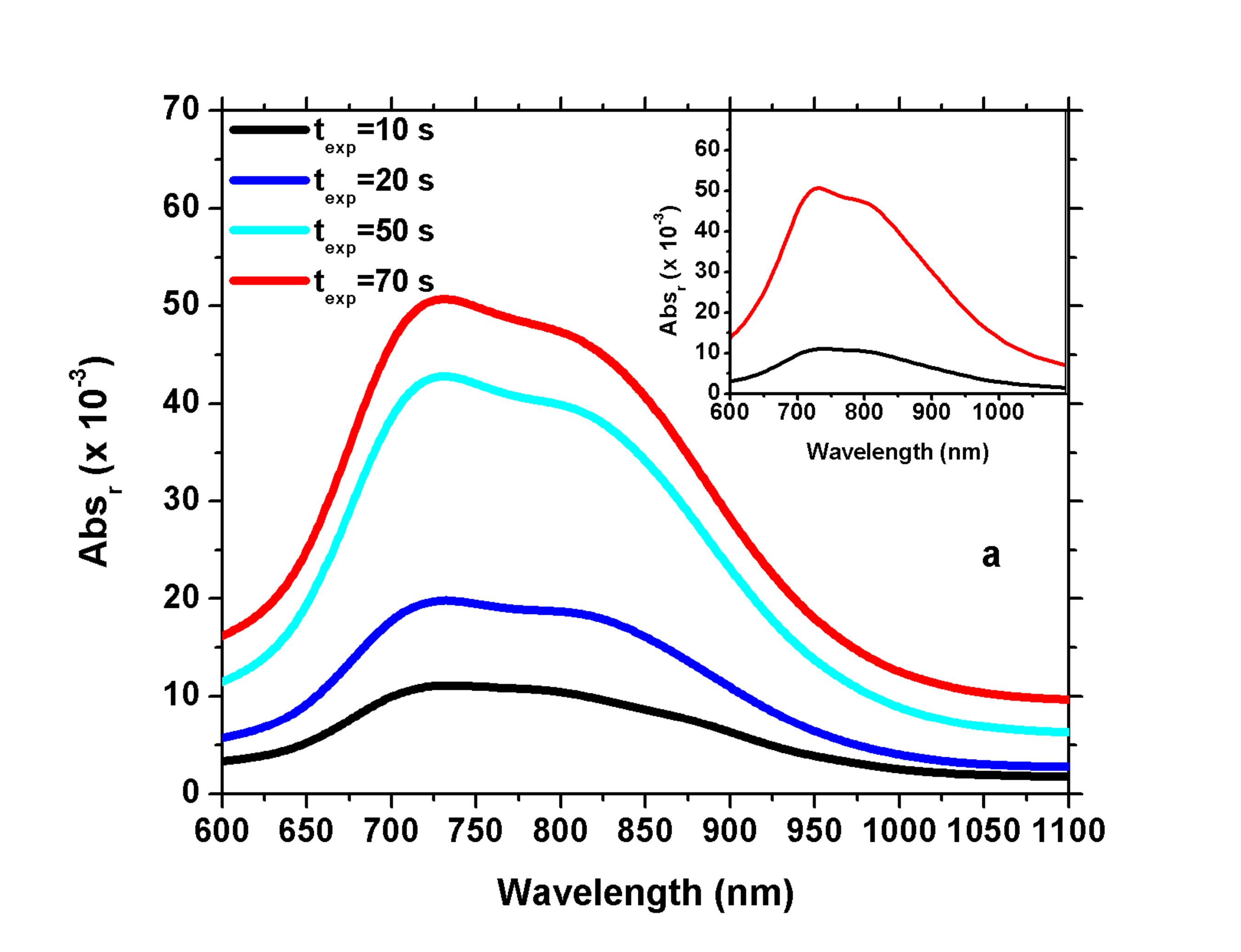}\\
\includegraphics[width=\columnwidth]{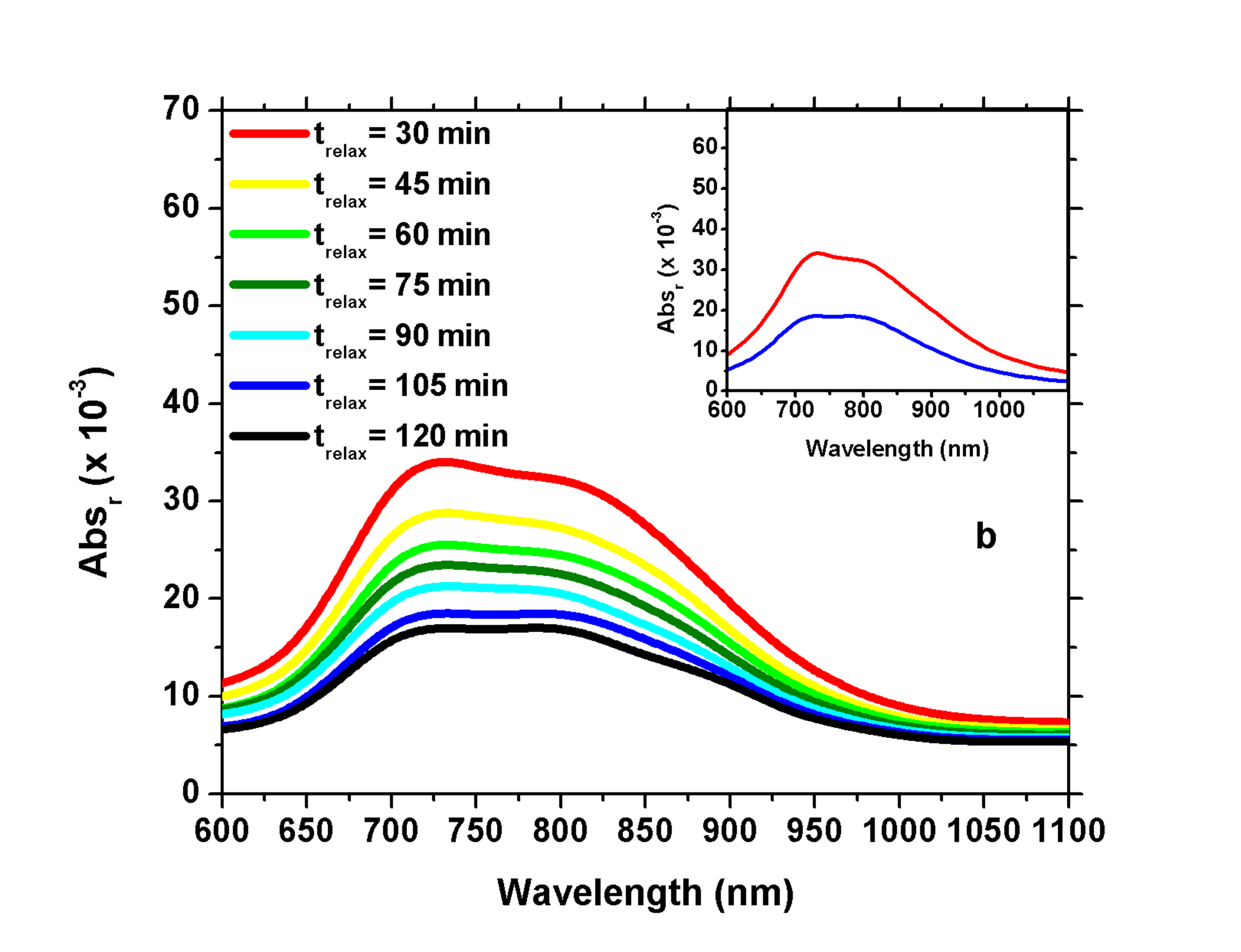}
\caption{\textbf{a}: Referenced sample absorbance after exposure to 5 mW/cm$^{2}$ at 405 nm. \textbf{b}: Referenced sample absorbance during relaxation in the dark after exposure to 5 mW/cm$^{2}$ at 405 nm for 70 s. Insets: resonances numerical fitting according to the Gans theory.}\label{fig:bluespectrum}
\end{center}
\end{figure}

By increasing the energy of the incident photons, the photodesorption rate rises \cite{burchianti2008}. As a consequence, the NPs' growth is more efficient with laser illumination at 405 nm: with a 70 times lower intensity and a 14 times shorter exposure with respect to the 532 nm case, a comparable absorbance variation is obtained upon illumination (Fig. \ref{fig:bluespectrum}.a). Apart from that, the absorption spectrum obtained after sample exposure to violet light exhibits some common features with the green case: two maxima at 730 nm and 830 nm appear. According to numerical fits following Eq. \ref{eqn:gans}, also the nanoparticles grown by the 405 nm laser light are 2 nm radius spheroids, mainly with an aspect ratio of 1.25 for oblates and 0.75 for prolates, respectively.

As in the previous case, the shape of the resonance band changes as the exposure time increases;  however, after 10 s of illumination, the prolates' peak is larger than the oblates' one. This implies that violet light forces the self-assembly of prolate nanoparticles more effectively since the beginning as a consequence of a deeper and almost instantaneous modification of the equilibrium interaction between the K atomic adsorbate and the substrate. Such different optical response with respect to the green case proves that the cluster growth is wavelength-dependent.

In the dark, the system relaxes back towards the initial condition, by light-grown clusters evaporation. Also in this case, then, the NPs' formation is completely reversible. Moreover, the absorption band continuously modifies its shape until the relative height of the oblates' peak exceeds the prolates'. Hence, the K nanoclusters' decay mechanisms depend on intrinsic factors, independent from the details of the illumination. 

\begin{figure}[h]
\begin{center}
\includegraphics[width=\columnwidth]{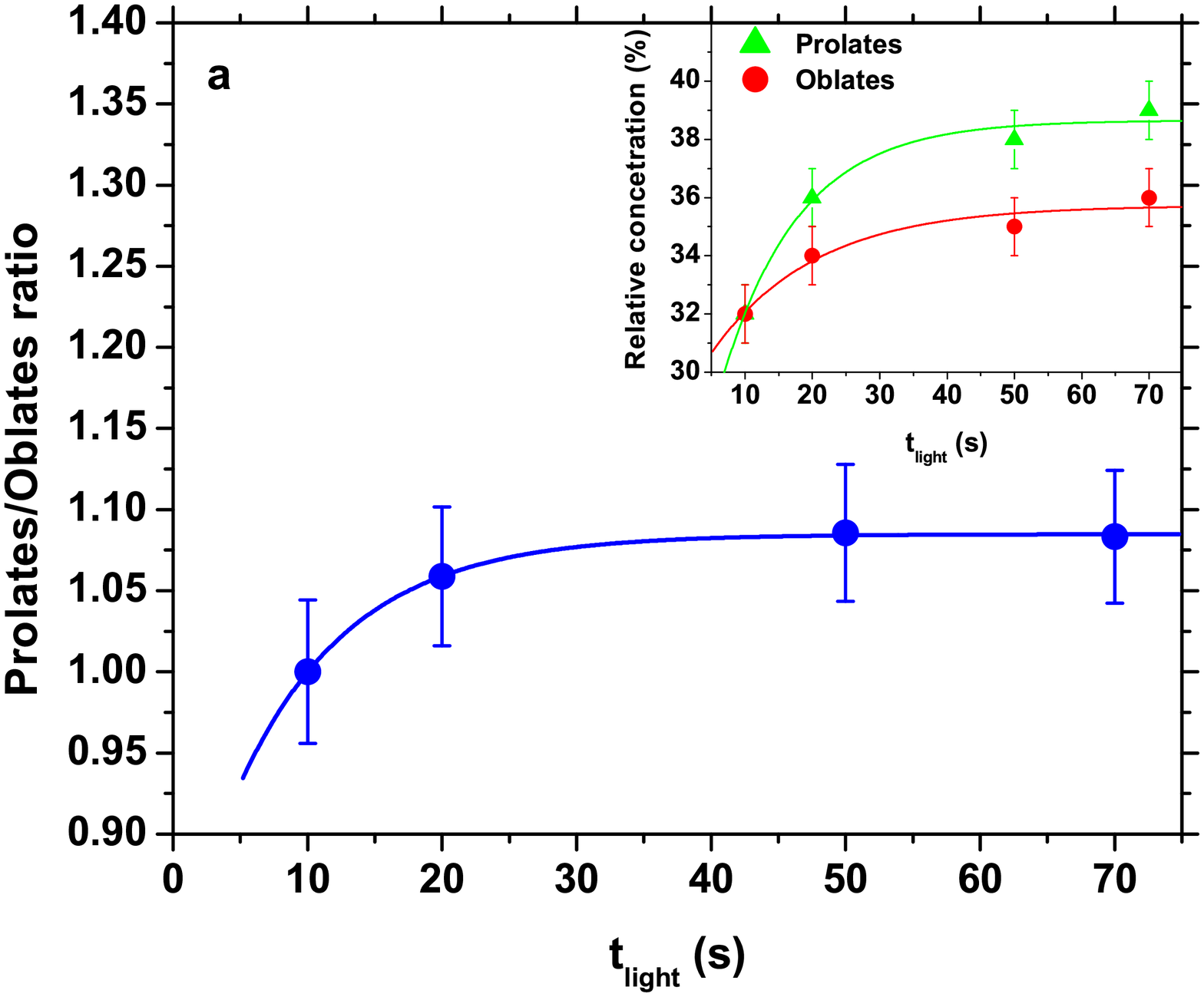}\\
\includegraphics[width=\columnwidth]{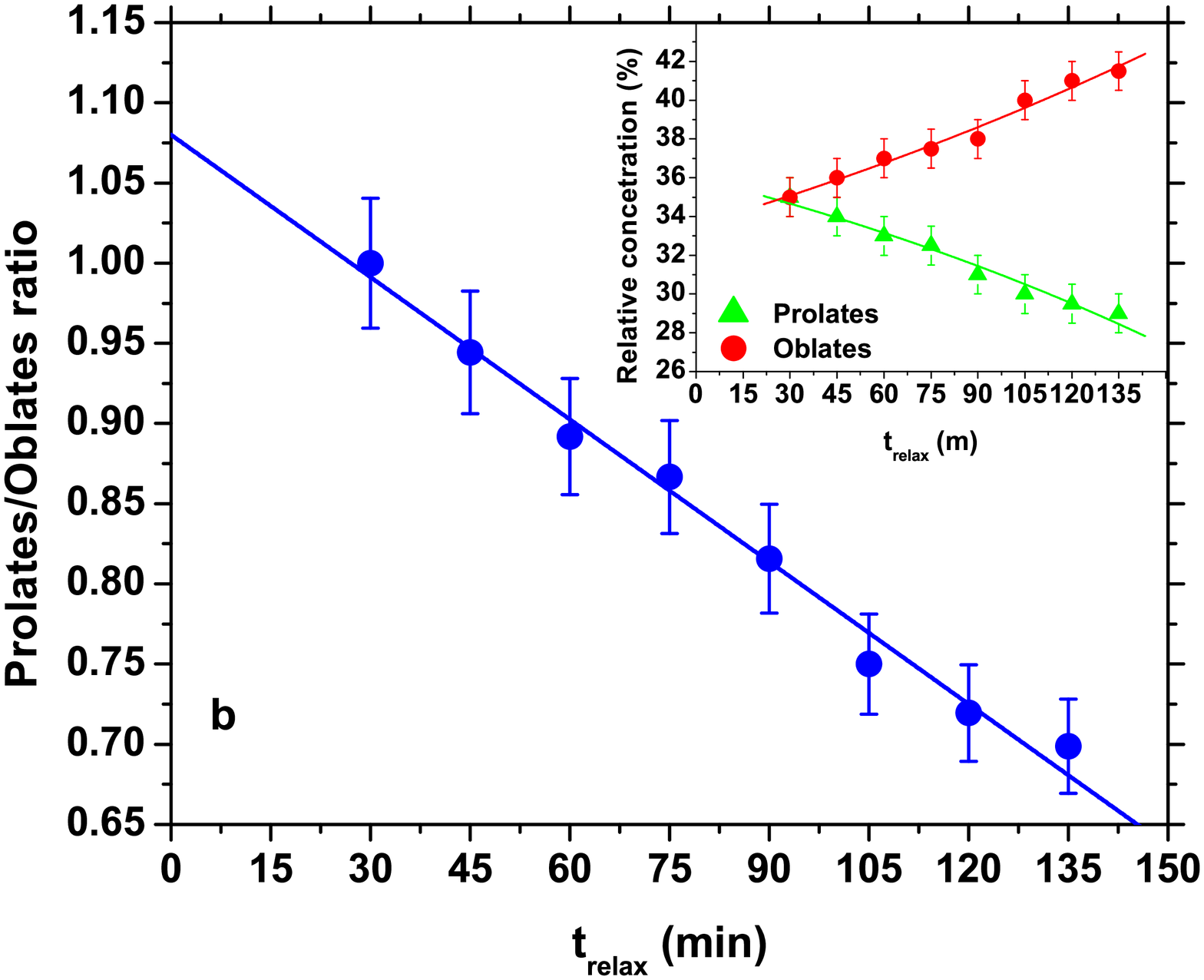} 
\caption{Ratio between the number of oblate and prolate NPs as a function of time for violet illumination according to the Gans theory fit results. \textbf{a}: 5 mW/cm$^{2}$ at 405 nm. \textbf{b}: relaxation in the dark. Lines are guides to the eye.}\label{fig:blueperc}
\end{center}
\end{figure}

These conclusions are confirmed by comparing the absorbance data with the extinction cross-section simulations (Eq. \ref{eqn:gans}). Fig. \ref{fig:blueperc}.a shows that, during illumination, the relative abundance of prolates is always equal or larger than the oblates, contrary to the 532 nm case. Hence, the violet radiation alters significantly the process of NPs' growth, quickly pushing the system towards a new regime. Relaxation is instead mainly influenced by the geometry instability of the light-grown K clusters. Indeed, in the violet case we also observed a restoration of the oblates' dominion over the prolates, in the context of an overall reduction of the NPs number. 

The light-induced modifications of the sample absorbance clearly indicate that illumination causes a reversible phase-transformation in our system. Indeed, as light hits the PG, the equilibrium inside the nanopores between atomic layers, oblate and prolate nanoparticles and vapour phase is altered.

In order to characterize the dynamics of this light-induced phenomenon, the flux of desorbed atoms in the cell volume and the change of the porous matrix transmission at 780 and 850 nm (TPBs) are measured as a function of time, during and after illumination. The TPBs' wavelengths are chosen in order to distinguish the contributions provided by prolate and oblate NPs to the sample's transparency. Under illumination the desorption probability of atoms, which are weakly adsorbed on the pore surface, increases. The atoms burst into the glass nanocavities can re-condense, either at surface defects or atomic layers. In the former case, they are trapped at the surface sites where, driven by light, metastable clusters are formed. In the latter case, instead, they are likely to photo-detach again and diffuse through the sample until they are ejected in the cell volume. Therefore, the evolution of $\delta(t)=(n(t)-n_{0})/n_{0}$, namely the relative variation of the K vapour density with respect to the equilibrium value $n_{0}$, is the result of the interplay of light-induced atomic desorption and reaction processes at the surface defects.

\begin{figure}[h]
\begin{center}
\includegraphics[width=\columnwidth]{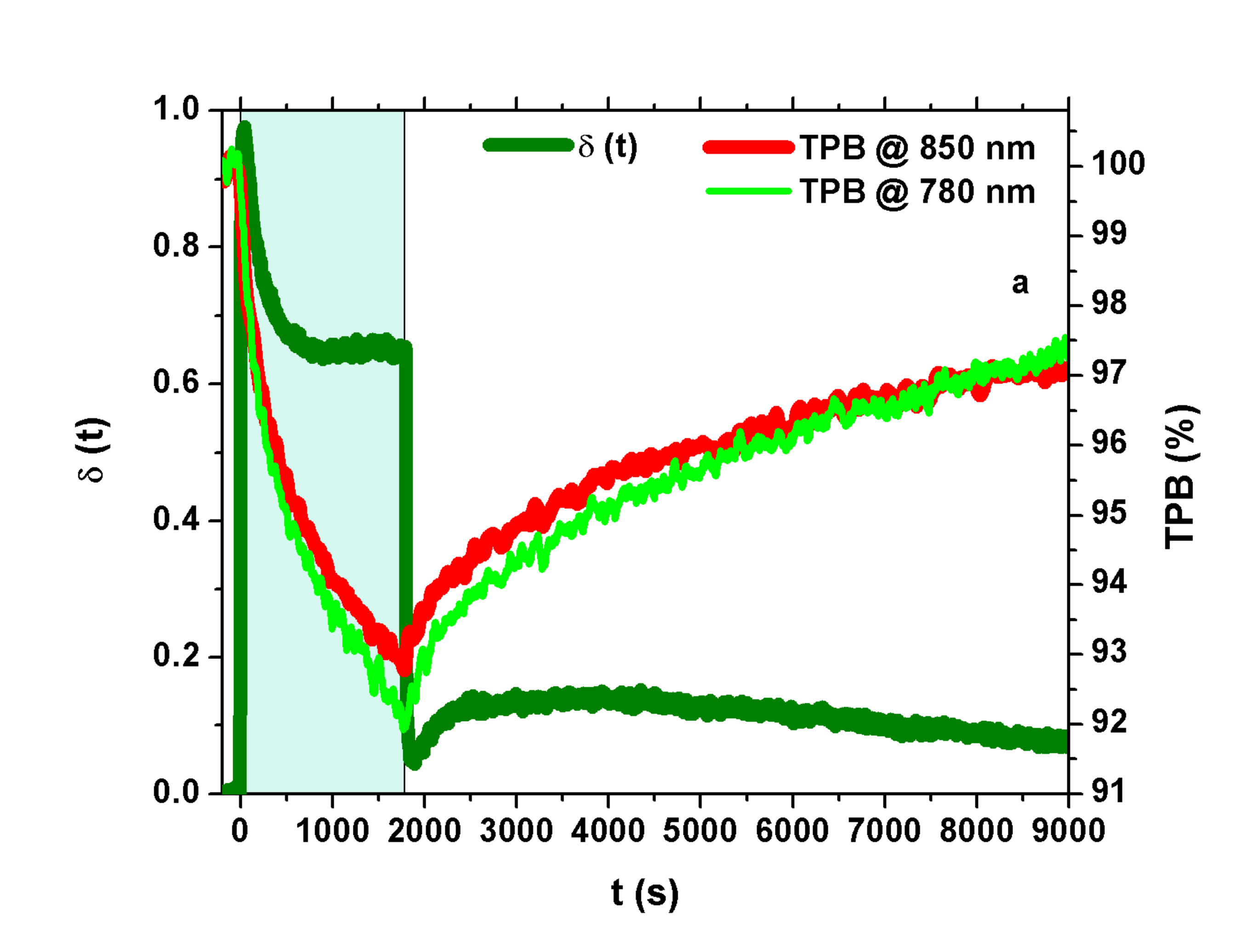}
\includegraphics[width=\columnwidth]{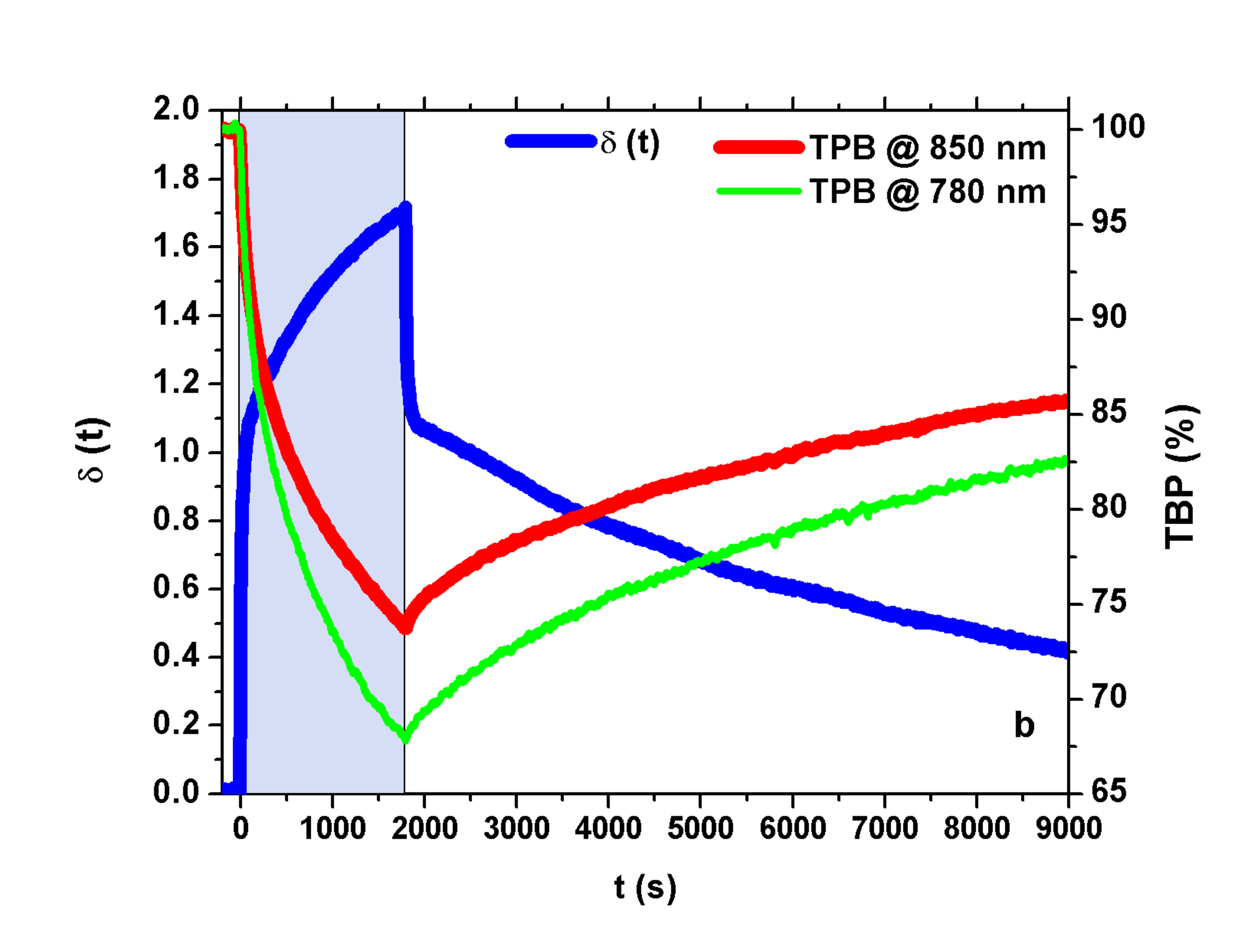}
\caption{$\delta(t)$ and PG transmission as a function of time. \textbf{a}: exposure to 40 mW/cm$^{2}$ at 532 nm.  \textbf{b}: exposure to 5 mW/cm$^{2}$ at 405 nm. Desorbing light is switched on at t=0 and then off at t=1800 s, as indicated by the shaded area.}\label{fig:dynamics}
\end{center}
\end{figure}

Since the average structural stability of the light-grown clusters is affected by the illumination wavelength and the exposure time, also the diffusion-reaction dynamics of the atomic flux through the nanopores result dependent on the details of the illumination. This clearly appears in the $\delta (t)$ and in the TPBs' signals shown in Fig. \ref{fig:dynamics}.a and Fig. \ref{fig:dynamics}.b for the green and violet illuminations, respectively.

In both cases, in fact, under exposure to laser light, the number of atoms desorbed in the cell volume increases and, at the same time, the sample transparency at 780 and 850 nm decreases. This confirms that light moves atoms from the pore surface both into the vapour phase and to the ``seed'' sites that act as nucleation centres for cluster growth. The features of this last process are consistent with the previous results of the absorbance analysis. Indeed, 405 nm light is more efficient in building NPs; moreover, the evolution of the two populations of spheroids is remarkably different for green and violet illumination.

In fact, in the green case, during the first tens of seconds, no difference between the levels of the TPBs is appreciable; this corresponds to the shortest illumination sequence in Fig. \ref{fig:greenspectrum}, when the relative height of the oblates' absorbance peak is slightly larger than the prolates'. After about t=250 s, however, the TPB at 780 nm decreases faster than the one at 850 nm. The system has now entered in the regime corresponding to the intermediate illuminations in Fig. \ref{fig:greenspectrum}: from now on, prolates are more effectively grown. The transition between the different regimes of NPs assembly is a smooth, continuous process: no evidence of a threshold is found.  

In the violet case, the TPBs are already significantly spread at t$>$10 s. This corresponds to data of Fig. \ref{fig:blueperc}.a after 10 s: the relative amount of prolates and oblates is the same in these conditions. After that, the prolate population progressively overwhelms the oblate one. 

In the dark, the TPBs, in both green and violet cases, progressively rise, consistently with the results of the absorbance analysis. It is worth noting that a sudden increase of the sample transparency is observed right after the interruption of illumination: highly unstable light-grown spheroids are present during illumination and, as soon as they are no longer compensated by new structures, they significantly affect the TPBs levels. At about t=2000 s, regardless the characteristics of the previous illumination, a second slower relaxation effect is onset and reestablishes the sample's equilibrium transparency. Remarkably, the slopes of the two TPBs are different: as already deduced from the shape evolution of the absorbance spectra, light-grown prolates are intrinsically less stable than oblates.

The differences observed in the TPBs' signals are also reflected in the behaviour of $\delta (t)$. This confirms that light drives the diffusion-reaction dynamics of the atomic flux in the porous matrix. At a given time, the value of the relative vapour density variation is in fact the result of the balance between the atomic desorbing flux from the sample and the escape fluxes towards the nanopores' walls and the metallic reservoir. In particular, in the case of the 532 nm illumination, $\delta(t)$ reaches its maximum value after a few seconds and then starts to drop because the desorption flux does not compensate the atomic losses anymore. This is due to a progressive depletion of weakly adsorbed atoms lying on the atomic layers as well as to the reaction events at the pore surface defects.

However, such decrement is slowed down: at t=850 s, $\delta(t)$ is almost stable and slightly increases. This can be attributed to a delayed flux of atoms produced by the evaporation of light-assembled unstable NPs, whose number is now relevant. Consistently, when the light is switched off, $\delta(t)$ abruptly drops towards the equilibrium value, but, as a consequence of the aforementioned fast evaporation of light-grown spheroids, the density increases again. For t$\geq$4000 s this contribution to the atomic flux becomes negligible due the progressive reduction of the number of NPs and the vapour density decreases again.

In the case of the 405 nm illumination, instead, after the sudden increase at t=0 s, the vapour density continues to grow with a slower trend. In fact, violet light since the beginning favours the formation of prolate NPs, whose fast evaporation provides an atomic flux which effectively compensates the pores' depletion. When the light is switched off, the density quickly drops down, but the source of atoms supplied by the large amount of prolates contrasts this reduction. A new regime, slower than the one observed after green illumination, is then imposed for the equilibrium recovery. We can thus deduce that the laser irradiation at 405 nm not only produces more unstable NPs, but also causes a more substantial alteration of the adsorbate/substrate properties, which affects the system evolution even during the relaxation in the dark.
 
In summary, according to the experimental results, light not only drives the atomic diffusion along the pores, by means of the light-enhanced desorption probability, but also induces surface diffusion at the surface defects. Indeed, we find that the shape of the formed aggregates depends on the light wavelength;  in particular, prolate NPs are more effectively grown by high energy desorbing photons. This implies that violet light increases also the surface mobility of strongly adsorbed atoms near the cluster boundary. This causes the growth of unstable structures characterised by a small surface contact. In this context, we remark that if the intensity of the 405 nm laser light is reduced, both the desorbing and the cluster growth rates decrease, however the main features of the atomic desorption flux and the TPBs' signals are not affected. On the other hand, by increasing the intensity of the 532 nm laser light, the behaviour observed in the violet case cannot be reproduced. In addition, for both illuminations, we observe a progressive shift towards a regime more favourable to the formation of prolate NPs. This effect can be attributed to an ongoing instability of the substrate-adsorbate complex,  caused by the exposure to light, which enhances the surface atomic mobility. As a consequence of all these process, also the atomic bulk diffusion, which is driven by the adsorption/desorption events along the nanochannels, exhibits modifications both during the illumination and the recovery in the dark.

In conclusion, our results show that, in nanoporous glass, the adsorption/desorption and nanoparticle formation processes are strongly influenced by low-power laser irradiation. In particular, we found that exposure to visible light has a double effect: on the one hand, it moves weakly adsorbed atoms from atomic layers to surface defects, where clusters are grown; on the other hand, it drives the atomic surface diffusion on the forming cluster.  By properly choosing the laser wavelength and the illumination time, we demonstrated that it is possible to control the relative concentration of oblate and prolate light-grown NPs. Furthermore, since the lifetime of the self-assembled clusters depends on their shape, we found a close relationship between the clusters' structural properties and the light-driven atomic flux along the nanochannels. 

This suggests the possibility to optically modify in ``real-time'' the overall atomic transport dynamics in porous media. Therefore, our all-optical approach for controlling the interactions between the adsorbate and the substrate provides an alternative tool for reversible self-assembly of  atomic nanostructures. It also represents a valuable playground  for testing reaction-diffusion models, that are recognised as promising assets for non-equilibrium complex materials \cite{mann2009, denisov2010}.

\begin{ack}
The authors are grateful to A. Bogi for the fruitful discussions. The authors acknowledge the cell manufacturing by M. Badalassi (CNR) and thank C. Stanghini and L. Stiaccini (University of Siena) for their technical assistance. 
\end{ack}

\end{document}